\documentclass[10pt, conference]{IEEEtran}
\IEEEoverridecommandlockouts

\usepackage{cite}

\usepackage{graphicx}
\usepackage{textcomp}
\usepackage{xcolor}
\usepackage{tikz}
\usepackage{amsmath}
\usepackage{filecontents}
\usepackage{amsmath,amssymb,amsfonts}
\usepackage{algorithm}
\usepackage{algorithmicx}
\usepackage{algpseudocode}
\usepackage{subcaption}
\usepackage{caption}
\usepackage{float}
\usepackage{verbatim}
\usepackage{soul}
\usepackage{multirow}
\usepackage{booktabs}
\usepackage{listings}
\usepackage{makecell}
\usepackage{url}
\usepackage{authblk}
\floatstyle{plaintop}
\restylefloat{table}
\usepackage[normalem]{ulem}
\def\BibTeX{{\rm B\kern-.05em{\sc i\kern-.025em b}\kern-.08em
    T\kern-.1667em\lower.7ex\hbox{E}\kern-.125emX}}
\begin{document}

\title{TensorLib: A Spatial Accelerator Generation Framework for Tensor Algebra
\thanks{Yun Liang is the corresponding author.}}
  

\author[1]{Liancheng Jia}
\author[1]{Zizhang Luo}
\author[1]{Liqiang Lu}
\author[1,2]{Yun Liang}
\affil[1]{Center for Energy-Efficient Computing and Applications, School of EECS, Peking University}
\affil[2]{Pengcheng laboratory, China}
\affil[ ]{
\{jlc, semiwaker, liqianglu, ericlyun\}@pku.edu.cn
        }
\date{}
\maketitle
\IEEEaftertitletext{\vspace{-1\baselineskip}}
\begin{abstract}

Tensor algebra finds applications in various domains, and these applications, especially when accelerated on spatial hardware accelerators, can deliver high performance and low power. Spatial hardware accelerator exhibits complex design space. Prior approaches based on manual implementation lead to low programming productivity, rendering thorough design space exploration impossible. In this paper, we propose TensorLib, a framework for generating spatial hardware accelerator for tensor algebra applications. TensorLib is motivated by the observation that, different dataflows share common hardware modules, which can be reused across different designs. To build such a framework, TensorLib first uses Space-Time Transformation to explore different dataflows, which can compactly represent the hardware dataflow using a simple transformation matrix. Next, we identify the common structures of different dataflows and build parameterized hardware module templates with Chisel. Our generation framework can select the needed hardware modules for each dataflow, connect the modules using a specified interconnection pattern, and automatically generate the complete hardware accelerator design. TensorLib remarkably improves the productivity for the development and optimization of spatial hardware architecture, providing a rich design space with trade-offs in performance, area, and power. Experiments show that TensorLib can automatically generate hardware designs with different dataflows and achieve 21\% performance improvement on FPGA compared to the state-of-the-arts.

\end{abstract}

\section{Introduction}

Tensor algebra is a prevalent tool and has been successfully applied to a broad range of applications such as machine learning, data analytics. Tensor algebra features different dimensions, sizes and computation patterns, which require special hardware acceleration. For example, the 2-D convolution is one of the most popular tensor operations in deep learning applications\cite{DBLP:conf/nips/KrizhevskySH12}. It involves a 4-D weight and 3-D input and requires to accumulate the partial sums in four dimensions\cite{DBLP:conf/nips/KrizhevskySH12}. MTTKRP is a widely used tensor operation for tensor factorization in recommendation systems, which takes one 3-D tensor and two matrices as inputs and generates a matrix. Due to the regular computation patterns, spatial hardware accelerators are commonly used for the acceleration of tensor algebra ~\cite{DBLP:conf/isca/JouppiYPPABBBBB17,DBLP:conf/isca/ChenES16,autosystolic,xiao20fcnn,xiao2021}.



\begin{figure}[t]
    \centering
    \includegraphics[width=\columnwidth]{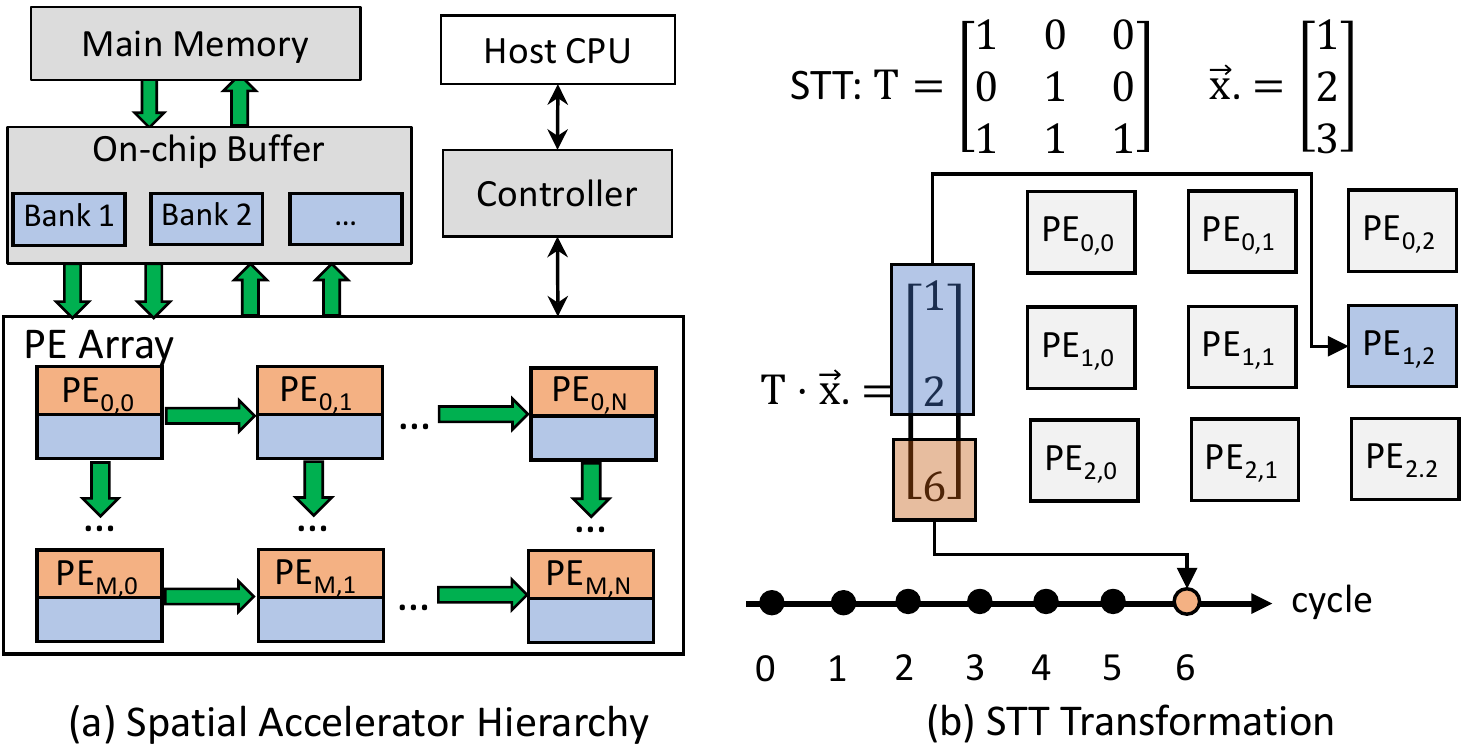}
    \caption{Spatial accelerator and Space-Time Transformation}\label{fig:systolicflow}
    \vspace{-1em}
\end{figure}

The majority of the spatial accelerator designs follow a hierarchical architecture\cite{lu2021tenet}. As shown in Figure~\ref{fig:systolicflow}, spatial accelerators usually consist of an array of homogeneous processing elements (PEs), an on-chip network that connects PEs together, a shared scratchpad buffer, and a system controller. The array of PEs can provide huge parallelism, and the connection between PEs can exploit different types of data reuse. While most spatial accelerators adopt the same hierarchical architecture, the actual implementation of each design can vary a lot. 
Among the various design parameters, hardware dataflow plays the most important role as it determines how the tensor is computed and communicated between PEs. There exists a sufficient large design space of dataflows for the spatial hardware accelerator design. Initially, dataflow is categorized by specifying the tensors that are reused temporarily inside each PE. For example, \cite{autosystolic} uses output stationary systolic array dataflow because the output tensor elements stay inside PE during execution. Similarly, \cite{DBLP:conf/isca/JouppiYPPABBBBB17} uses weight stationary dataflow and \cite{DBLP:conf/isca/ChenES16} uses row stationary dataflow.

The complex structure of spatial accelerator and its large design space leads to low productivity, rendering thorough design space exploration impossible.  To improve the programming productivity, HLS tools have been used for accelerator designs, which supports hardware generation with software-style programming\cite{autosystolic,cong2018polysa}. Some recent works also design DSLs or other notations to represent the dataflow and hardware architecture of spatial accelerators~\cite{t2stensor,lai2020susy,heterocl,lu2021tenet}. The user can use DSL to express the architecture and the compiler generates hardware code in HLS. However, the auto-generated HLS code is hard to optimize, resulting in low performance. Recent advances in hardware programming introduce highly parameterized and modular design principles based on Scala and Python~\cite{chisel}. The high-level language can be equally expressive as Verilog which supports cycle-level RTL description, but they also support functional and object-oriented programming. The high-level programming features enable a large variety of hardware instances to be generated with the same hardware template, which is extremely useful for spatial accelerators with a variety of dataflows.

In this paper, we propose TensorLib, a framework for generating spatial accelerator for tensor algebra. We use Space-Time Transformation (STT)~\cite{DBLP:conf/asap/BaltusA93} as a means of expression to represent the hardware dataflow.  STT maps loop instances to hardware spatially (coordinates in the PE array) and temporally (timestamp of execution) and it can cover the complete design space of spatial dataflows with a linear transformation. By inferring the pattern of data reuse using STT, we can directly represent various dataflows used in spatial accelerators including unicast, systolic, multicast, stationary, etc. 


 


We observe that different hardware dataflows can share common hardware modules at each level of the hierarchical structure. In other words, a new dataflow can be derived from an existing dataflow by only modifying certain parts. Based on these findings, we develop a few basic hardware component templates for each level of hierarchy using Chisel~\cite{chisel}, facilitating hardware reuse of different dataflows. When a dataflow is specified by STT, TensorLib can automatically select the templates and connect them together to build the hardware architecture. Some recent HLS frameworks also apply STT~\cite{cong2018polysa,lai2020susy,DBLP:journals/micro/JiaLWL20}, but they are limited in both generability and performance. First, their proposals mainly target systolic architectures which do not cover the complete space of dataflows, and only support limited applications. Second, the tensor algorithm is tightly coupled with architecture in their frameworks, making it hard to optimize. We overcome these challenges by providing a thorough dataflow analysis based on STT and building highly parameterized hardware module templates. Overall, TensorLib makes the following contributions,

\begin{itemize}
    \item We propose TensorLib, a framework to automatically generate various hardware dataflow implementation of spatial accelerators for tensor algebra applications.
    \item We develop a formal representation of spatial hardware dataflows using space-time transformation which can cover a comprehensive design space of dataflows.
    \item We integrate the STT-based dataflow analysis with Chisel-based hardware template to build a accelerator generator with rich design space, high productivity and good performance.
\end{itemize}
Experiment shows that TensorLib is able to generate a large number of dataflow architectures for various tensor applications. Compared to the state-of-the-art, Tensorlib achieves 21\% throughput improvement on FPGA for matrix multiplication benchmark, and supports more dataflows and applications, which remarkably improves both performance and generality. The source code of Tensorlib is available at \url{https://www.github.com/kirliavc/tensorlib}.

\section{Space-Time Transformation Background}
Space-time transformation (STT) is a linear transformation that maps the tensor algebra to hardware from a spatial and temporal perspective~\cite{DBLP:conf/asap/BaltusA93}. STT can be used for tensor algebra whose computation can be described in perfect nested loops. The PE array can be viewed as a hypercube, and the execution of hardware can be identified as a space vector and a time scalar indicating where and when the computation takes place. STT transforms a point in the loop nest to the space-time vector in hardware execution using a matrix multiplication operation. For example, given a loop iteration in the loop nest $\vec x = [i, j, ...]^T$ and a transformation matrix $T$, the execution space and time can be calculated as follows,
\begin{equation}
   \left[\begin{matrix} \vec p \\ t \end{matrix}\right]=Tx
    \label{STTdef}
\end{equation}
where space vector $\vec p$ means the PE coordinates inside the PE array and time scalar $t$ means the time step of execution. Figure \ref{fig:systolicflow} (b) shows an example using matrix multiplication $C[i,j]+=A[i,k]\times B[k,j]$ as the target tensor algebra. There are three iterator variables $i$, $j$ and $k$. When $i=1,j=2,k=3$, using equation \ref{STTdef} we can get $\left[\begin{matrix} \vec p \\ t \end{matrix}\right]=(1,2,6)^T$, which means $A[1,3]\times B[3,2]$ takes place at PE $(1,2)$ at the sixth cycle. Since a PE can only perform one operation per cycle, matrix \textbf{T} must be a full-ranked matrix so that there is an one-to-one mapping between iterator space and space-time space.
\vspace{-0.5em}
\section{Framework}
\begin{figure}[t]
    \centering
    \includegraphics[width=0.45\textwidth]{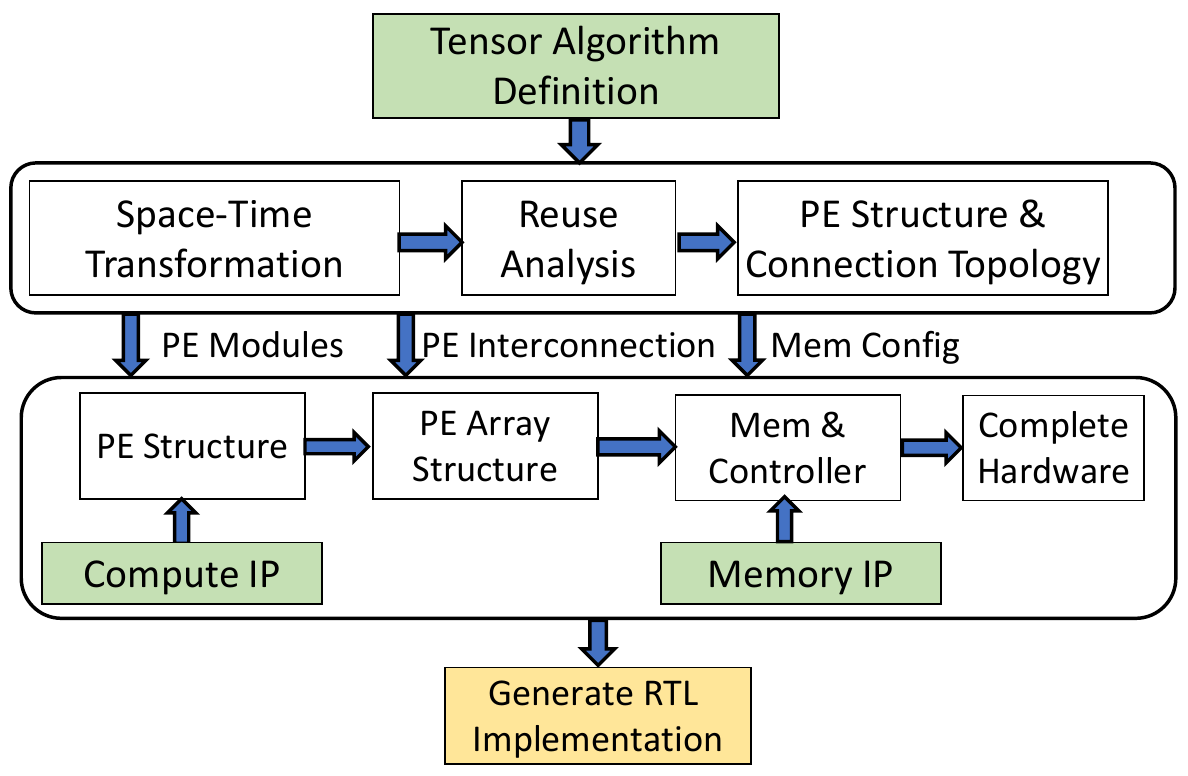}
    \caption{The overview of TensorLib}\label{fig:workflow}
    \vspace{-1em}
\end{figure}

Figure \ref{fig:workflow} presents the overview of TensorLib framework. The workflow can be divided into two steps: dataflow generation and hardware implementation generation. For the dataflow generation, TensorLib uses the tensor algebra described by a nested loop and an STT matrix ($T$ in Equation~(\ref{STTdef})) as input. The framework first maps the nested loop into the spatial hardware PE array using STT. After that, by analyzing the recurrence of the same tensor element in different space-time vectors, TensorLib determines the reuse pattern and dataflow type of each tensor. The computation involves multiple tensors and STT generates different dataflows for each tensor. We categorize the dataflow for each tensor into five types: Unicast, Stationary, Systolic, Multicast (Reduction Tree for output) and 2D-reuse. Details are discussed in Section IV.

The second step is hardware generation. The framework generates the 3-level hierarchy of spatial accelerators in a bottom-up manner. It first uses the dataflow type of each tensor to select the internal modules of PE and connect them with the computation IP to form the PE structure. Next, it connects the PEs together with the generated patterns to form the PE array. Finally, it generates the memory modules with access pattern and external memory IPs, and the controller which provides control signals for both PE and memory ports. Details are discussed in section V.



\section{Dataflow Generation}
To represent a dataflow, it is critical to capture the movement of each tensor. Given a loop iteration $\vec x$, the tensor index accessed by $\vec x$ can be expressed as $\vec I = A\vec x$, where $A$ is the corresponding access matrix. The mapping between tensor index and loop iteration index can be 1-to-N so that the same tensor element is accessed by multiple loop iterations. By transforming the loop iterations to the space-time domain, the reuse of the same tensor element forms the \textit{reuse hyperplane} in space-time space, which determines how the tensor moves (e.g. dataflow). By replacing the $\vec x$ in Equation \ref{STTdef} with accessing matrix, we have,
\begin{equation}
    AT^{-1}\left[\begin{matrix}\vec p \\ t \end{matrix}\right]=\vec I
    \label{hyperplane}
\end{equation}

Based on this, we can build a reuse subspace composed of all the space-time points that reuse the same tensor index. Next, based on the rank of the reuse subspace, we divide the reuse shape into three cases as shown in Table I. Firstly, if the rank is 0, the hyperplane is actually a single point, which means the tensor element only appears once in the whole computation without any reuse. The unicast dataflow requires large on-chip bandwidth since every PE reads data from the on-chip memory independently.
\begingroup
\begin{table}
\small
\centering
\caption{Dataflow analysis with STT} \label{tab:motivation}
\begin{tabular}{|c|c|c|c|} 
\toprule[1pt]

\makecell[c]{Subspace \\ Dimension} & Shape & \makecell[c]{Space-Time \\ Reuse Space} & \makecell[c]{Tensor \\ Dataflow} \\

\hline
0           &         - &  \begin{minipage}[c][12mm][t]{0.1mm}%
\end{minipage} \makecell{\includegraphics[scale=0.5]{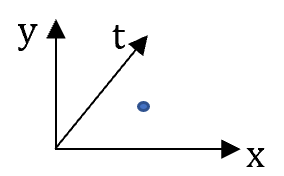}}                &     Unicast\\       
\hline

\multirow{3}{*}{1}  & $d\vec p=0, dt\neq 0$ &  \begin{minipage}[c][12mm][t]{0.1mm}\end{minipage}
\makecell{ \includegraphics[scale=0.5]{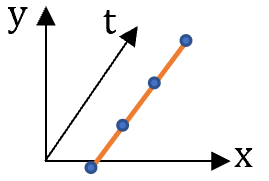}}                &     Stationary\\ 
\cline{2-4}
& $d\vec p\neq 0, dt\neq 0$ &\begin{minipage}[c][10mm][t]{0.1mm}\end{minipage} \makecell{\includegraphics[scale=0.5]{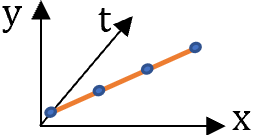}}  & Systolic\\
\cline{2-4}
& $d\vec p\neq 0, dt= 0$ &\begin{minipage}[c][11mm][t]{0.4mm}\end{minipage} \makecell{\includegraphics[scale=0.5]{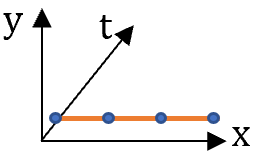}}  & Multicast \\
\hline
\multirow{3}{*}{2}  & \makecell[c]{t-axis \\Vertical} &  \begin{minipage}[c][10mm][t]{0.1mm}\end{minipage}
\makecell{ \includegraphics[scale=0.5]{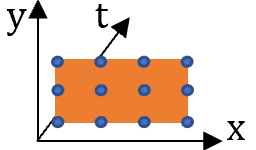}}                &     Broadcast\\ 
\cline{2-4}
& \makecell[c]{t-axis \\ Parallel} &\begin{minipage}[c][12mm][t]{0.1mm}\end{minipage} \makecell{\includegraphics[scale=0.5]{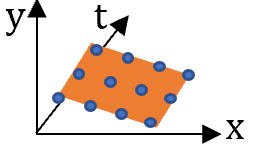}}  & \makecell[c]{Multicast \\ \& Stationary}\\
\cline{2-4}
& \makecell[c]{t-axis \\ Intersect} &\begin{minipage}[c][12mm][t]{0.08mm}\end{minipage} \makecell{\includegraphics[scale=0.5]{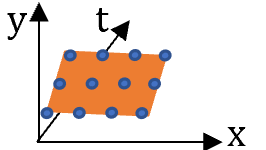}}  & \makecell[c]{Systolic \\ \& Multicast} \\
\bottomrule[1pt]
\end{tabular}
\vspace{-1em}
\end{table}
\endgroup

\begin{figure*}[t]
    \centering
    \includegraphics[width=\textwidth]{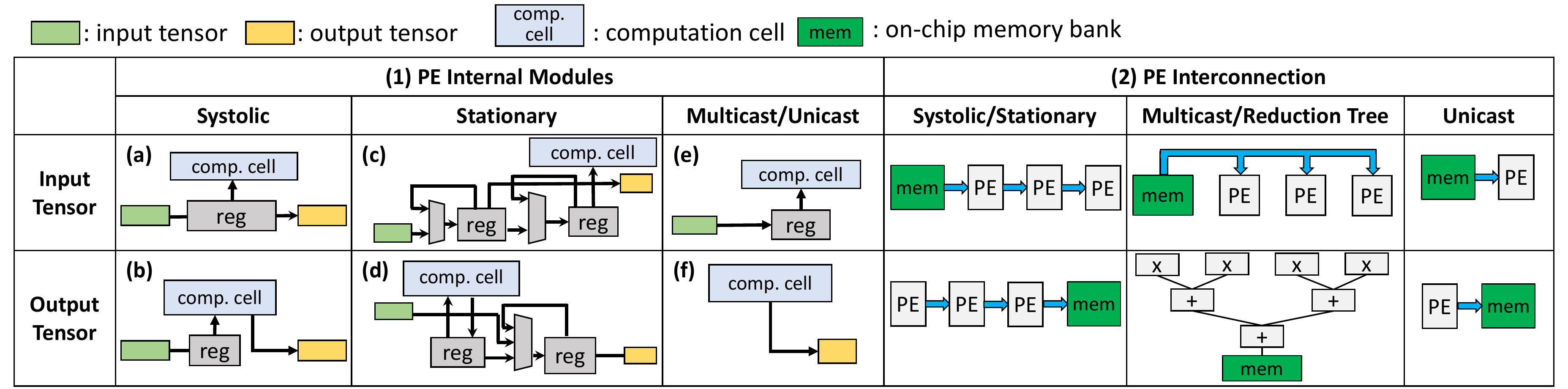}
    \caption{The PE internal modules and interconnection modules for different dataflows}\label{fig:impldataflow}
    \vspace{-1em}
\end{figure*}

If the rank equals to 1, the hyperplane is a straight line, and the dataflow is determined by the direction of the line. We can calculate $d\vec p$ and $dt$ easily using the base vector of the hyperplane, as showed by Equation \ref{dxdydt}, where $Eig$ is the first eigenvector, $E$ is the identity matrix, $(AT^{-1})^-$ is the pseudoinverse of $AT^{-1}$. 

\begin{equation}
    \left[\begin{matrix} d\vec p \\ dt \end{matrix}\right] = Eig\left(E-(AT^{-1})^-(AT^{-1})\right)
    \label{dxdydt}
\end{equation}

There are totally three sub-cases:
\begin{itemize}
    \item $d\vec p=\vec 0, dt\neq 0$. The same tensor element appears in the same PE, but at different time steps, which means that the tensor is stationary and stays inside one PE.
    \item $d\vec p\neq \vec 0, dt\neq 0.$ The tensor element appears in different PEs and at different time steps. This corresponds to systolic dataflow where data are delayed for one cycle before being sent out to other PE.
    \item $d\vec p\neq \vec 0, dt=0.$ The tensor element appears in different PEs but at same time step. The same data must be transformed to different PEs at the same time, which corresponds to a multicast dataflow. If it is a output tensor, the dataflow indicates that different partial results is generated simultaneously by different PEs, which requires a reduction tree to generate final results.
\end{itemize}

Finally, if the rank is 2, the reuse subspace is a 2D plane. In this case, the shape of reuse space is categorized based on the relationship with time axis. There are totally three sub-cases:
\begin{itemize}
    \item Vertical to t-axis. The same tensor element is broadcasted to all the PEs in the array at the same cycle. 
    \item Parallel to t-axis. The tensor element is firstly broadcasted to a group of PEs, and the element keeps stationary in the PE during the execution stage.
    \item Intersect with t-axis. The tensor element is firstly broadcasted to a group of registers, and then traverses between PE in a systolic style.
\end{itemize}
The PE array dataflow and interconnection pattern of 2D reuse subspace is similar to the 1D case because both of them are formed with 1D reuse patterns (systolic, multicast and stationary). Finally, typical tensor algebras such as 2D Convolution contain more than three loop nests. For a 2D PE array, we need to select three loops (2D space + 1 time) from the loop nest to map to space-time space. The remaining loops are executed sequentially which doesn't influence PE dataflow. When PE and memory sizes are determined, the loops are performed tiling to fit the hardware resources.

Using the example in Figure \ref{fig:systolicflow} (b), the accessing matrix for $A[i,k]$ is  $\left[\begin{matrix} 1 & 0 & 0 \\ 0 & 0 &1 \end{matrix}\right]$. With Equation (3), we can find that the dimension of reuse subspace is 1 and $\left[\begin{matrix} d\vec p \\ dt \end{matrix}\right]=(0,1,1)^T$, which means tensor A uses the systolic dataflow with vertical direction. 
\begin{figure}[t]
    \centering
    \includegraphics[width=0.9\columnwidth]{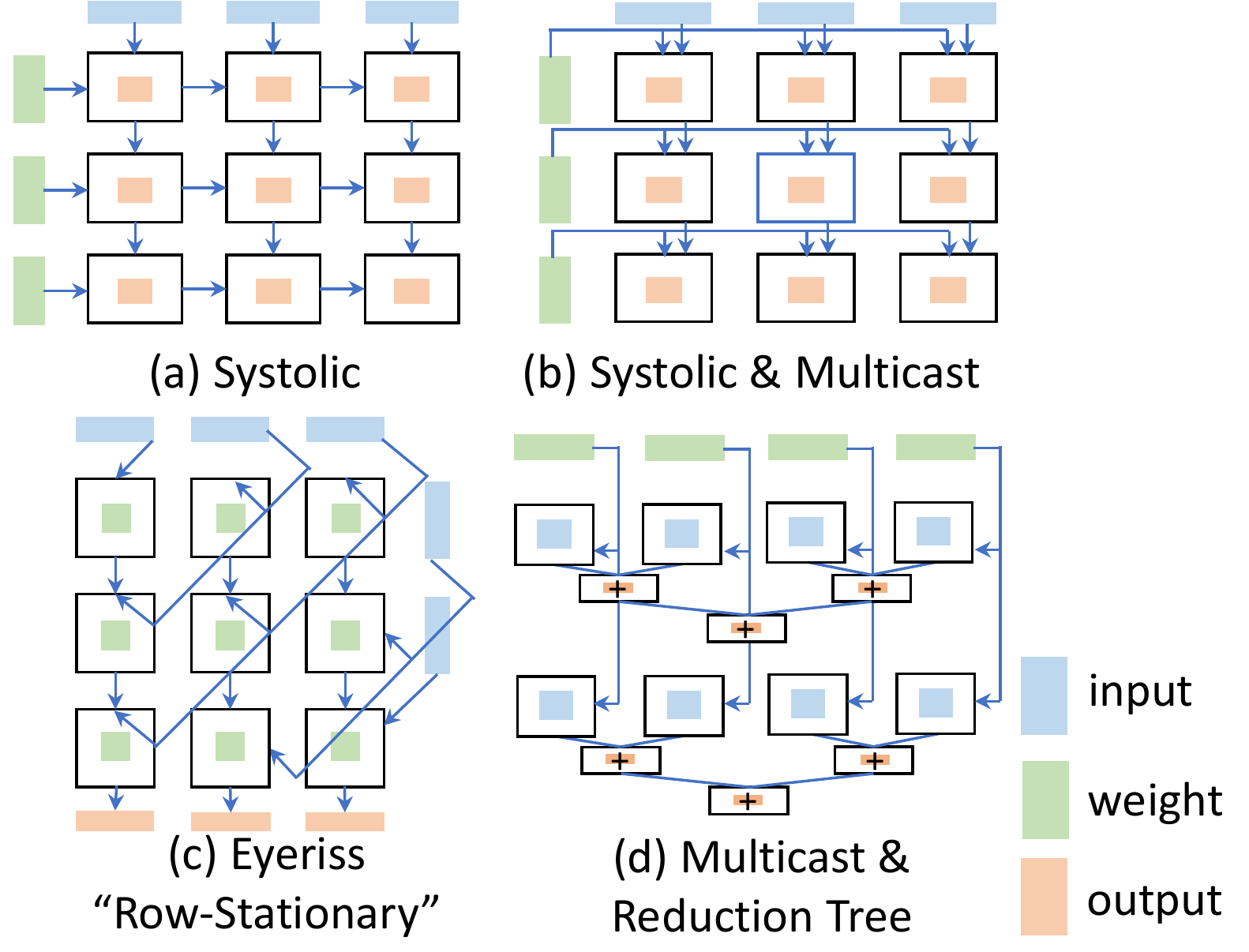}
    \caption{Examples of PE interconnection for different dataflows}\label{fig:interpe}
    \vspace{-1em}
\end{figure}
\section{Hardware Generation}

After dataflow generation, the next step is to implement the architecture with the specific dataflow. The architecture of spatial accelerators can be separated into three levels: PE structure, PE-Array interconnection, on-chip shared buffer and controller. In this section, we discuss how to generate each level of hardware architecture for different dataflows. 




\subsection{PE Generation}
The PE structure consists of the computation cell and the internal modules that connect the cell with I/O ports. The computation cells are manually implemented IPs. The difference in dataflows mainly lies in the internal modules that connect the I/O ports and the computation cells. Since each tensor can use different dataflow and the internal modules for each tensor do not connect to each other, they can be implemented independently. As shown in Table 1, the dataflow of each tensor inside PE has three types: (a) systolic, (b) stationary and (c) multicast or reduction tree. Each tensor can be either input or output of the algorithm. The implementation is different for input and output tensor because the output tensor needs to read results from the computation cell, so there are 6 conditions in total. In TensorLib framework, we build the implementation of PE internal modules for each condition, and the circuit diagram is presented in Figure \ref{fig:impldataflow}. 

\begin{itemize}
    \item Modules (a) and (b) are designed for systolic dataflows. Tensor elements always transfer to their neighboring PEs every cycle. The difference is that the output data is generated from the computation cell and the input data is transferred to the next PE directly. 
    \item Modules (c) and (d) are designed for stationary tensors. Tensor elements stay inside PE during execution, but the data needs to be updated when the execution stage ends. It uses a double-buffer structure to enable the parallelism of computation and data communication. In (d), one register is used to update the result of the current stage, and the other is used to transfer the results of the last stage.
    \item Modules (e) and (f) are designed for multicast and unicast dataflows, where data are received and passed directly.
\end{itemize}

Different PE dataflow can be constructed by selecting the corresponding modules. For example, output stationary~\cite{autosystolic} dataflow contains two modules (a) (systolic, input) and one module (d) (stationary, output). Weight stationary~\cite{DBLP:conf/isca/JouppiYPPABBBBB17} dataflow contains one (a), one (b) (systolic, output) and one (c) (stationary, input). Eyeriss~\cite{DBLP:conf/isca/ChenES16} dataflow uses one multicast and one stationary module for input, and systolic module for output. The selected components are connected with the PE body and computation cell components to generate the complete PE structure.

\subsection{Interconnection and Memory Generation}


Different dataflows also require different connection topologies between PEs.  Figure \ref{fig:impldataflow} (2) shows the PE interconnection patterns for different dataflows. The systolic and stationary dataflows connect adjacent PEs together. The direction of systolic dataflow interconnection is determined by the reuse vector $(d\vec p, dt)$. The output of $PE(x, y)$ connects to $PE(x+dx, y+dy)$ after delaying $dt$ cycle. For multicast input dataflow, the same input data is transferred from the on-chip buffer to PEs in a row directly at the same cycle. The output multicast dataflow is implemented with a reduction tree to perform a reduction on the output of PEs. For unicast dataflow, different PEs are totally independent and connect to the on-chip memory bank directly. The next step is on-chip memory generation. Each group of PEs that reuse the same tensor indexes is assigned with a particular memory bank. TensorLib automatically generates the memory module and connects to the PE array with the pattern shown in Figure \ref{fig:impldataflow} (2). We design a flexible template for the memory module which supports different load/store patterns. 


Figure \ref{fig:interpe} shows four examples of PE interconnection patterns based on different types of dataflow for GEMM (the connection of stationary dataflow isn't shown). In systolic dataflow (a), PEs are connected one by one, and data is transferred to the adjacent PE every cycle in a fixed direction. For multicast input dataflow (b), tensor elements are read from shared on-chip bank and broadcasted to every PEs simultaneously. Eyeriss~\cite{DBLP:conf/isca/ChenES16} dataflow uses diagonal connection for input multicast dataflow as shown in (c). Finally, for multicast output dataflow, different PEs generate their partial sums at the same cycle, and they are connected with reduction trees to generate the final results as shown in (d).

\begin{table}[t]
    \small
    \centering
        \caption{Evaluated Tensor Algebras}
     \resizebox{\columnwidth}{!}{
    \begin{tabular}{|c|c|}
        \hline
        Name & Formula\\
        \hline
GEMM & $C[m, n] += A[m, k] \times B[n, k]$\\
Batched-GEMV & $C[m, n] += A[m, k, n] \times B[m, k]$\\
Conv2D & $C[k, y, x] += A[c, y+p, x+q] \times B[k,c,p,q]$\\ 
Depthwise-Conv & $C[k, y, x] += A[k, y+p, x+q] \times B[k, p, q]$\\
MTTKRP & $D[i,j]+=A[i,k,l]\times B[k,j] \times C[l,j]$\\ 
TTMc &$ D[i,j,k] += A[i,l,m]\times B[l,j]\times C[m,k]$\\
        \hline
    \end{tabular}
    }
\vspace{-1em}
    \label{dataflowtype}
\end{table}

\vspace{-0.4em}
\section{Experimental Evaluation} 
Given the dataflow specified by STT, TensorLib generates the hardware implemented in Chisel~\cite{chisel}. The generated Chisel code is compiled to Verilog and then is synthesized on both ASIC and FPGA platforms. For ASIC evaluation, we use Synopsys DC compiler with 55nm UMC 1P8F technology for synthesis, and Synopsys VCS for simulation. For FPGA evaluation, we use Xilinx VU9P with 6840 DSPs and 2160 BRAMs. We use Xilinx Vivado 2019.1 software for FPGA synthesis. For FPGA floating-point multiplication, we use Xilinx's Floating-Point IP and integrate it into Chisel implementation as a BlackBox module. 

We evaluate six tensor applications: GEMM, Batched-GEMV, Conv2D (2 layers from ResNet), Depthwise-Conv2D, MTTKRP and TTMc. The formula of each tensor algebra is shown in Table 1. For each tensor, we use S, T, M, U, B to refer to systolic, stationary, multicast (reduction tree), unicast, 2-D reuse space (implemented with multicast \& systolic) dataflow, respectively. To represent a dataflow using STT, we first need to select three loop iterators from the loop nest. Therefore, in the following, we name the dataflow of the hardware using the selected loop iterators and the dataflow of each tensor. For example, XPQ-MMT refers to selecting X, P, Q loops to perform STT transformation, and use multicast dataflow for A and B, and Stationary for C, respectively. For Conv2D, XYP-MMT is the dataflow with multicast connection, KCX-SST and KCX-STS are well-known output-stationary and weight-stationary systolic array dataflows\cite{autosystolic,DBLP:journals/corr/abs-1911-09925,DBLP:conf/isca/JouppiYPPABBBBB17}, and XYP-MST is similar as ShiDiannao's dataflow\cite{DBLP:conf/isca/DuFCILLFCT15}. 
\begin{figure}[t]
    \centering
    \includegraphics[width=\columnwidth]{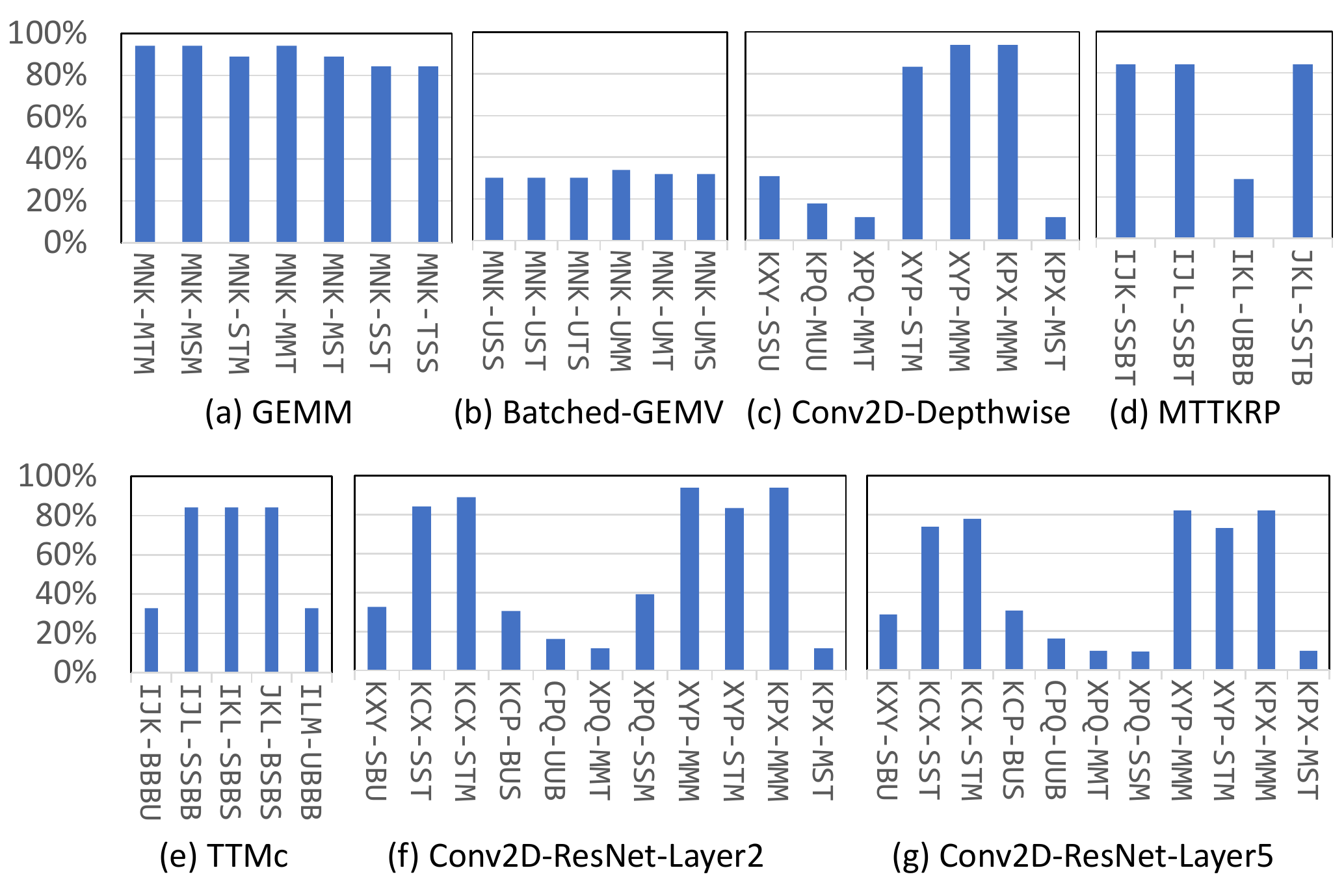}
    \caption{Normalized performance of different dataflows for each tensor algebra}\label{fig:utilization}
    \vspace{-1em}
\end{figure}

\subsection{Performance Results}
We evaluate the execution cycles of different dataflows with simulation. We set the size of PE array to $16\times 16$, it runs under 320MHz frequency with 32GB/s on-chip bandwidth between PE array and scratchpad memory. Figure \ref{fig:utilization} presents the normalized performance of a few representative dataflows measured by execution cycles compared with peak performance (full PE array utilization). As shown, different dataflows vary greatly in performance. 

\vspace{-0.1em}
For GEMM benchmark, the performance of multicast dataflows (MTM) is better than systolic dataflow (STS) because multicast dataflows have a smaller pipeline overhead than systolic array. But systolic array is preferred in hardware because of the lower interconnection cost and better frequency. For MTTKRP and TTMc, the unicast dataflows (e.g. IKL-UBBB and IJK-BBBU) perform worse than others because unicast dataflows require all PEs to transfer data with on-chip memory simultaneously and bandwidth becomes insufficient. Batched-GEMV can only use unicast dataflow because the tensor A is only accessed once and cannot be reused during computation. For Conv2D, some selected loops contain a small iteration range (e.g. Conv2D kernel size P and Q can be 3), leading to low utilization of PEs. For example, XYP-SMM and KPX-TMM have 1/16 idle PEs since the range of $p$ is 3 and only 15 out of 16 rows of PE are used. The performance of ResNet-Layer5 is even lower because X and Y loops are also small (x=y=7). For the KPX-MST dataflows used in Conv2D, although a PE is assigned with workloads, it becomes idle in some cycles because of communication delay. When the execution cycle is small, the communication delay can be larger than computation, which greatly hurts performance. For Conv2D workloads, selecting KCX iterations can deliver better performance because it becomes standard GEMM operation with large loop bounds. However, for Depthwise-Conv, a large reduction dimension doesn't exist, so regular Conv2D dataflows cannot be applied to Depthwise-Conv. The KPX-MMM and XYP-MMM dataflow perform better than other dataflows for Depthwise-Conv.

\subsection{Power and Area Evaluation}
\begin{figure}[t]
    \centering
    \includegraphics[width=\columnwidth]{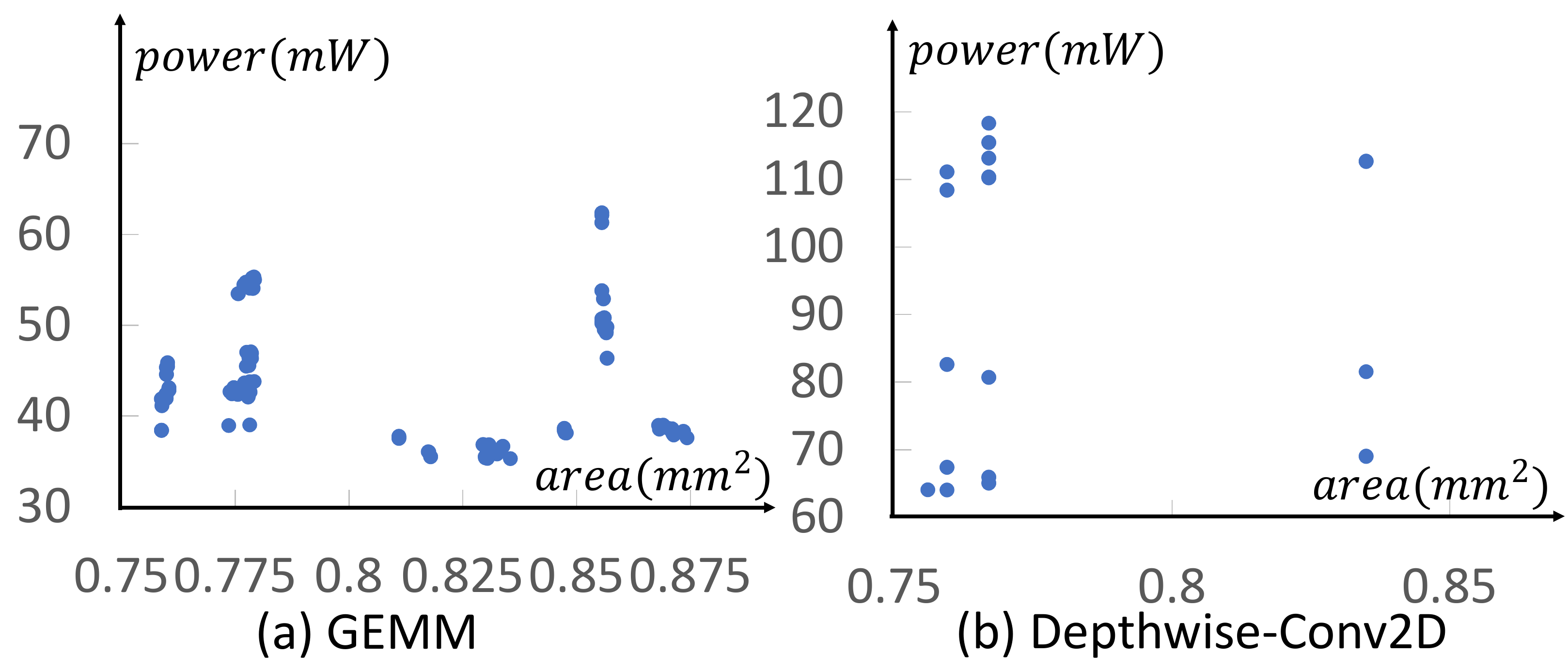}
    \caption{Power and area result of different dataflow designs}\label{fig:areapower}
    \vspace{-1em}
\end{figure}

Here, we evaluate a large design space of dataflows for GEMM and Depthwise-Conv2D in a $16\times 16$ PE array for INT16 datatype. We set the target frequency to 320MHz for ASIC synthesis. Figure \ref{fig:areapower} gives the area and energy performance of different dataflow architectures generated by TensorLib. Each point in the graph refers to one dataflow, and there are totally 148 points for GEMM and 33 points for depthwise-Conv2D. Experiments show that dataflow choice has a larger impact on energy consumption than area. The energy variation of GEMM range from 35mW to 63mW, which shows 1.8X difference, while the area has only 1.16X difference. Compared with other dataflows, dataflow with two multicast input (MMT, MMS) consumes more energy. However, reduction tree output dataflow doesn't cost too much energy, although they have similar STT-level representation. Dataflows with stationary tensor also consume more area and energy because of the control signals for stationary data.

\begin{table}[t]
    \small
    \centering
        \caption{FPGA Performance Comparison on MM workload}
    \resizebox{\columnwidth}{!}{
    \begin{tabular}{|c|c|c|c|c|c|c|}
    \hline
    & \multicolumn{2}{c|}{Susy\cite{lai2020susy}} & \multicolumn{2}{c|}{PolySA\cite{cong2018polysa}}  & \multicolumn{2}{c|}{TensorLib} \\
        
        \hline
Device & \multicolumn{2}{c|}{Arria-10}  & \multicolumn{2}{c|}{VU9P}  & \multicolumn{2}{c|}{VU9P} \\
\hline

      Workload   & MM & Conv & MM & Conv & MM & Conv\\
LUT &40\%& 35\%& 49\%& 49\% &68\%&73\%\\ 
DSP &93\%& 84\%& 89\%& 89\% & 75\%&75\%\\
BRAM &32\%& 30\%& 89\%& 71\%& 51\%&73\%\\ 
MHz &202& 220& 229& 229& 263&245\\
Gop/s &547& 551& 555& 548& 673&626\\
        \hline
    \end{tabular}
    }
\vspace{-1em}
    \label{dataflowtype}
\end{table}

\vspace{-0.7em}
\subsection{Comparison with Prior works}
\vspace{-0.3em}
We also compare TensorLib with PolySA \cite{cong2018polysa} and Susy \cite{lai2020susy}. PolySA and Susy can generate the hardware implementation of systolic arrays with a polyhedral model or STT. However, their designs are limited in performance and supported algorithms. Both Susy and PolySA only support systolic array dataflow, and they fail to generate hardware for algorithms that don't fit well in systolic architecture, such as Depthwise convolution. We synthesize our implementation with systolic array (KCX-STS) dataflow on MM and Conv2D workloads with FP32 datatype. The size of PE array in our FPGA implementation is $10\times 16$ and the vectorization degree in each PE is 8. The result is shown in Table III. Our synthesis result achieves 673 Gop/s throughput and 263 MHz frequency, which obtains 21\% throughput and 15\% frequency improvement compared with the state-of-the-art generators. 

Recently, AutoSA\cite{autosa} extends PolySA with I/O optimization and more tensor algorithms. It also uses Autobridge\cite{DBLP:conf/fpga/GuoC0LQUZC21} to improve the frequency by assigning each hardware module to one particular FPGA slot and minimizing the slot-crossing cost. Inspired by this, we also manually optimize the physical placement using Vivado toolchain. This will improve the frequency of MM design to 328 MHz on VU9P.


\vspace{-0.3em}
\section{Conclusion}
In this paper, we propose TensorLib, a framework for generating spatial accelerator for tensor algebra. TensorLib uses STT to analyze the tensor reuse behavior to generate the dataflow type and communication direction for each tensor. We build reusable hardware module templates for each dataflow and Tensorlib automatically select hardware modules to construct PE structure, PE interconnection, on-chip memory and controller for complete spatial accelerator. Experiment results show that TensorLib can generate various accelerators for tensor applications with different dataflows, and achieve better performance than state-of-the-art generators on FPGA.

\section*{Acknowledgment}

This work was supported in part by the Beijing Natural Science Foundation (No. JQ19014) , Beijing Academy of Artificial Intelligence (BAAI), and Key-Area Research and Development Program of Guangdong Province (No. 2019B010155002).

\bibliographystyle{plain}

\bibliography{ref}

\end{document}